# Titan's Atomic and Molecular Nitrogen Tori

H.T. Smith[a], R.E. Johnson[a], V.I. Shematovich[b]

[a]*Materials Science and Engineering, University of Virginia, Charlottesville, VA 22904 USA*
[b]*Institute of Astronomy, RAS, Moscow, RUSSIA*

## Abstract

Shematovich et al. (2003) recently showed plasma induced sputtering in Titan's atmosphere is a source of neutral nitrogen in Saturn's magnetosphere comparable to the photo-dissociation source. These sources form a toroidal nitrogen cloud roughly centered at Titan's orbital radius but gravitationally bound to Saturn. Once ionized, these particles contribute to Saturn's plasma. When Titan is inside Saturn's magnetopause, newly formed ions can diffuse inward becoming inner magnetospheric energetic nitrogen where they can sputter and be implanted into icy satellite surfaces. Our 3-D simulation produces the first consistent Titan generated N and $N_2$ neutral clouds; solar UV radiation and magnetospheric plasma subject these particles to dissociation and ionization. The cloud morphologies and associated nitrogen plasma source rates are predicted in anticipation of Cassini data. Since the amount of molecular nitrogen ejected from Titan by photo-dissociation is small, molecular nitrogen ions detection by Cassini will be an indicator of atmospheric sputtering.

## 1. Introduction

Beyond Saturn's five inner icy satellites lies Titan, its largest satellite. Pioneer 11, Voyager 1 & 2, Earth based observations and more recently, Hubble Space Telescope (HST) observations contributed much of our limited knowledge about the magnetosphere between Saturn and Titan. In the vicinity of the icy satellites, neutral atoms and molecules dominate the magnetosphere with only about 10% of these particles present as ions (Richardson 1998). However, we know little about the magnetosphere composition at Titan's orbit. Because Saturn's magnetic field rotates on an axis approximately aligned with planetary rotation, the thermal plasma efficiently interacts with the neutrals at average relative velocities varying with distance from the planet. Since Titan lacks a magnetic field, these plasma particles can also penetrate its dense $N_2$ atmosphere and eject neutrals into the magnetosphere, a process often referred to as atmospheric sputtering (Johnson 1994). Initial estimates from Voyager observations placed the Titan generated N source rate at 3 x $10^{26}$ N/s (Strobel and Shemansky 1982). Later, Strobel et al (1992) estimated that electron impact and photo-dissociation produced less than or equal to $10^{25}$ N/s. Using a 1-D Monte Carlo simulation and the pick-up ions flux from Brecht et al. (2000), Shematovich et. al. (2003) calculated an atmospheric sputtering source rate at approximately 4 x $10^{25}$ N/s, which is comparable to the photo- and electron impact dissociation source. Michael et al (2004) using a 3-D Monte Carlo model with improved cross sections obtained a slightly higher loss rate. In spite of the considerable uncertainties in these calculations, it is possible to conclude Titan provides a significant source of both atomic and molecular nitrogen forming a neutral torus in Saturn's magnetosphere. Electron impact, charge exchange and UV photons subsequently ionize these neutrals. Some of these ions diffuse inward toward Saturn providing a hot nitrogen source for implantation into the surfaces of the icy satellites (Sittler et al. 2004). Therefore, Cassini's plasma instruments will also search for nitrogen containing molecules ejected from the icy satellites surfaces.

We describe here a 3-D Monte Carlo simulation that calculates the density of the N and $N_2$ neutral clouds. We optimized the model to provide efficient execution over many years. Principal outputs are the neutral N and $N_2$ cloud topography and the spatial distribution of the freshly produced atomic and molecular nitrogen ions. The UV instrument and the energetic neutral imaging on Cassini might detect these neutrals and the Cassini Plasma Science instrument (CAPS) will directly detect freshly ionized nitrogen. Our model can readily accommodate new data as it becomes available. Although the model does not assume azimuthal symmetry, until Cassini plasma data is available we will use azimuthally symmetric ionization rates.

## 2. Model Description

Titan loses atmosphere through Jeans' escape; photo- and electron-impact dissociation; ionization and sweeping, and atmospheric sputtering. Whereas ionization and sweeping is a direct source of molecular ions for Saturn's magnetosphere, the other processes produce neutrals. Jean's escape primarily yields molecular hydrogen, dissociation produces atomic N and H, and atmospheric sputtering produces atomic and molecular neutrals, primarily N and $N_2$. Here we model the neutral nitrogen atoms and molecules ejected from Titan.

Our model has two main components. The "satellite model" replicates the environment close to Titan and the "Saturn model" reproduces the torus around Saturn. The satellite model produces high-resolution results within the Hill sphere and provides input to the "Saturn model", which generates particle densities for the entire neutral cloud. This separation will allow us to make the best use of the expected Cassini data at Titan and in the torus. We designed the model for rapid execution over relatively large time periods and we account for neutrals reabsorbed by Titan.

The simulations are initiated using nitrogen source fluxes from Titan (Michael et. al., 2004; Shematovich et al., 2003), ejected by ion and photon bombardment of its upper atmosphere. These fluxes also account for re-impact of pick-up ions. In the satellite model, N and $N_2$ are ejected in a Titan centered inertial reference frame. Although the model is 3-D, we use isotropic ejection from Titan until the plasma flow onto the exobase is better defined. We compare these to results using the standard atmospheric sputtering distribution (Johnson 1994):

$$F(E) = \frac{U}{(E+U)^2} \qquad (1)$$

where U is the escape energy from Titan's exobase (~0.3 eV for N and ~0.6 eV for $N_2$).

We characterize the nitrogen atoms and molecules escaping from Titan using the N and $N_2$ energy spectra in Fig.1 in combination with the source rates in Table 1. Michael et al. (2004) generated the pick-up ion source-using energetic $N_2^+$ to represent molecular pick-up ions of similar mass ($C_2H_5^+$, $H_2CN^+$, $HCN^+$, etc.). We used the photon-induced source from Shematovich et al. (2003). The model ejects N and $N_2$ particles from Titan's exobase and propagates their trajectories until they exceed Titan's Hill Sphere. At this point the satellite model transfers particles to the Saturn model. These particle trajectories are then propagated within the Saturn model until they impact Saturn, a satellite or the main rings, escape from Saturn or become ionized.

## 3. Results

Particle lifetimes and the initial energy distribution of particles escaping from Titan primarily determine neutral cloud densities. We used the same set of photo and electron impact cross sections used in Shematovich et. al. (2003) and charge exchange cross-sections are from our database (http://www.people.virginia.edu/rej/cross_section). We then applied these cross sections to two radial distributions of the plasma in Saturn's magnetosphere. The solid lines shown in Figs. 2a and 2b apply when Titan lies within Saturn's magnetosphere, representing lifetimes derived from Richardson and Sittler (1990) and Voyager data. Fig. 2a also shows the average lifetime used by Barbosa (1987) at Titan's orbit. When we conservatively extrapolated the lifetimes beyond 14 Rs, where Rs is the radius of Saturn, then photon-induced lifetimes dominate at Titan's orbit (as shown by the dotted lines in Figs. 2a and 2b). This gives an approximate lower bound to the ionization and dissociation rates and can roughly represent periods when Titan is outside of the magnetosphere.

Fig. 3 shows the N and $N_2$ densities based on the solid line lifetimes in Figs. 2a & 2b and the energy distributions in Fig. 1. The N cloud includes dissociated $N_2$ products as well as directly ejected N. This combination of short lifetimes and energetic nitrogen results in the lowest densities. Approximately 67% of the N and 47% of the $N_2$ escape beyond 50 Saturn radii and, therefore, do not contribute significantly to the nitrogen cloud's morphology. In addition, Titan reabsorbs a few percent.

To examine the initial energy's effect on cloud density distribution, we also used the standard energy distribution in Fig. 1 and the longer lifetimes (dotted lines) in Figs. 2a and 2b to create a rough upper bound to the nitrogen densities. Fig. 4 shows the resulting neutral cloud densities. These are noticeably increased as only 38% of the N and 34% of the $N_2$ escape beyond 50 Saturn radii.

These simulations give a neutral cloud with a significantly different morphology from the simple model of Barbosa (1987). Barbosa's estimated cloud extends inward to 8 Saturn Radii (Rs) while ours extends several Rs closer to Saturn and his cloud thickness is 8-16 Rs while ours is much less than 8 Rs. Additionally, Barbosa estimates an average nitrogen density of $6/cm^3$, but our average density is lower. Although he assumed a much larger vertical scale, his assumed source rate is over an order of magnitude larger, leading to higher densities. Ip (1992) also predicted higher densities (3-20 $/cm^3$) near Titan. He also assumed an order of magnitude higher source rate. In both of the simulations presented here, the nitrogen clouds extend inward towards Saturn and, therefore, are a source of fresh nitrogen ions in inner magnetosphere as well as in the outer magnetosphere. Fig.5 gives the source rate of fresh atomic and molecular nitrogen ions for the clouds in Fig. 3. Nitrogen is ionized faster in the inner magnetosphere so in this model the $N^+$ source rate peaks in this region despite the smaller neutral particle density.

Finally, our results show in the vicinity of Titan's orbital radius, the neutral nitrogen density is larger than the Voyager-based ion density as is the case in the inner magnetosphere (Richardson et al. 1998). However, both the nitrogen ion densities and our model for the neutrals are based on a number of assumptions and extrapolations which we will update as Cassini data becomes available.

## 4. Conclusions

In this paper we calculated the neutral tori at Saturn generated by N and $N_2$ escaping from Titan. The cloud density and morphology, and the spatial distribution of the nitrogen ions produced are given in anticipation of Cassini. These results can be scaled, as better Titan source rates are determined. The tori are centered on Titan and extend into the inner magnetosphere, with most of the density remaining close to the orbital plane. Our initial results indicate that neutral nitrogen may dominate the nitrogen ion population near Titan's orbital radius. The initial nitrogen energy distribution noticeably affects the cloud morphology and density. Relatively small increases in this energy distribution can cause many more particles to escape from Saturn since Titan is more than 20 Rs from Saturn in a region of relatively weak gravitational attraction. At this distance, the total escape energy from Saturn is only 4.3 eV for N and 8.6 eV for $N_2$. Therefore, Cassini measurements of the morphology will help constrain models of the plasma interaction with Titan.

Since Titan is not always within Saturn's magnetopause, we used ionization and dissociation rates that roughly bracket the times when it is inside and outside the magnetosphere. However, care must be taken, because even when Titan lies outside of the magnetopause on the dayside, it will re-enter the magnetosphere before reaching the night side. Because each Titan orbit is approximately sixteen days but the lifetimes are such that steady state requires a number of years of simulation, one would need to average these cases for the fraction of the time spent outside of the magnetopause. When Cassini data is available, we will use 3-D ionization rates to generate a correct time-varying model that allows for azimuthal asymmetry and varies over the solar cycle. We have also given ion source rates for one case. It is seen, surprisingly, that there is a small secondary peak in the nitrogen ion formation rate at a smaller distance from Titan caused by centrifugal confinement and higher ionization rates in this region. These ions will have lower average energies than inwardly diffusing nitrogen ions.

Although the $H_2$ source rate is higher than either N or $N_2$, the heavier nitrogen ions are of interest to us because of their potential impact on inner magnetospheric objects. The nitrogen ions that diffuse inward become energized and can be implanted in and sputter these moons (Sittler et al. 2004), ultimately driving nitrogen chemistry (Delitsky and Lane 2002). Because atmospheric sputtering is a significant source of $N_2$ as well as N, but photo-dissociation is not, detection of freshly produced $N_2^+$ ions in Titan's torus by the Cassini plasma science instrument (CAPS) can determine the relative importance of atmospheric sputtering and the nature of the plasma/atmosphere interaction.

**Acknowledgements**
We appreciate the positive remarks by both reviewers. This work is supported by NASA's Planetary Atmospheres Program, the NASA Graduate Student Research Program, the Virginia Space Grant Consortium Graduate Research Fellowship program and the CAPS Cassini instrument team.
.

Table 1. Source rates for nitrogen escaping from Titan for all modeled processes

| **Ejection Process** | **Source Rate (particles / second)** |
|---|---|
| N from $N^{+(1)}$ | 8.85E+24 |
| N from $N_2^{+(1)}$ | 1.15E+25 |
| N2 from $N^{+(1)}$ | 1.68E+24 |
| N2 from $N_2^{+(1)}$ | 4.21E+24 |
| N from hv $^{(2)}$ | 9.30E+24 |
| **Total** | **3.55E+25** |

(1) Michael et al. 2004; (2) Shematovich et al. (2003)

## Figure Captions

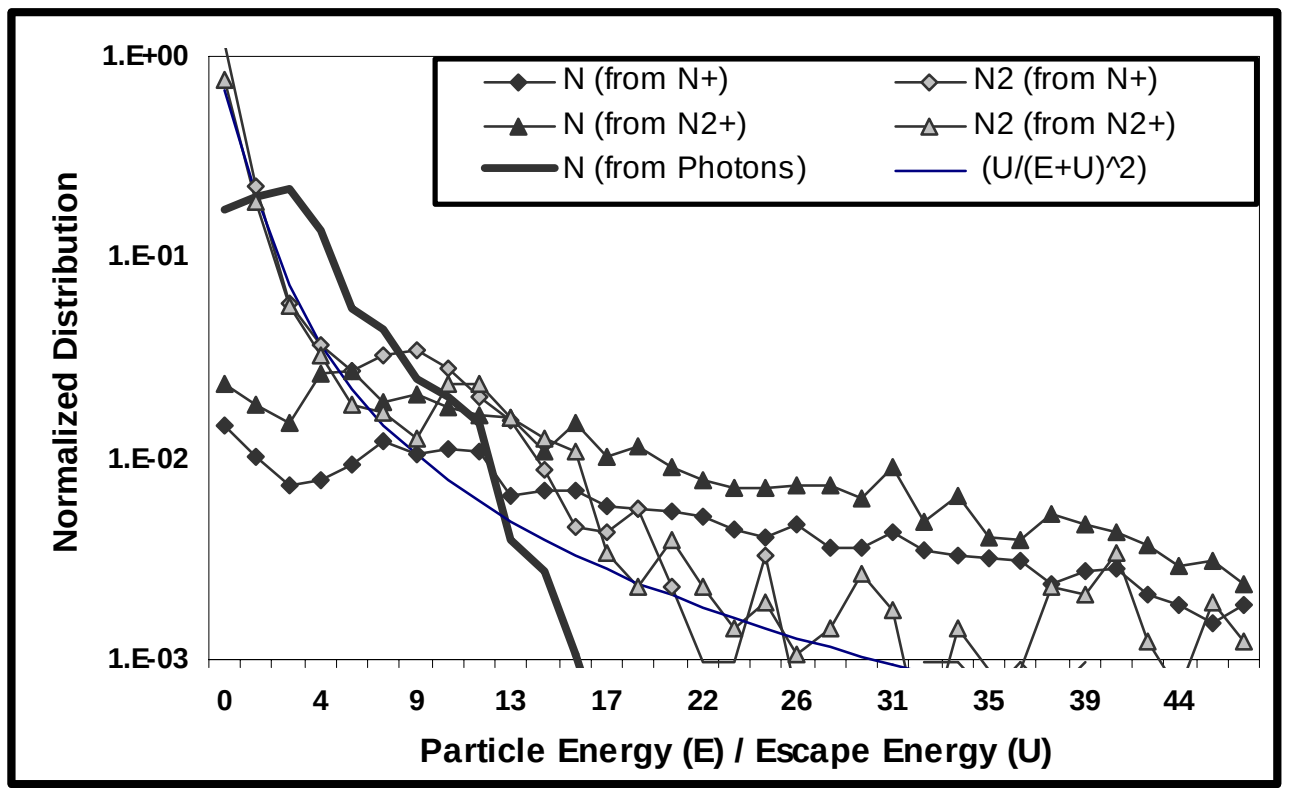

Figure 1. Normalized energy distributions for atomic and molecular nitrogen escaping from Titan vs. E/U (U=0.34eV for N and 0.68 eV for $N_2$) (Shematovich et al. 2003; Michael et al. 2004). Thin solid line: standard atmospheric sputtering distribution (Johnson 1990).

**(a)** **(b)**

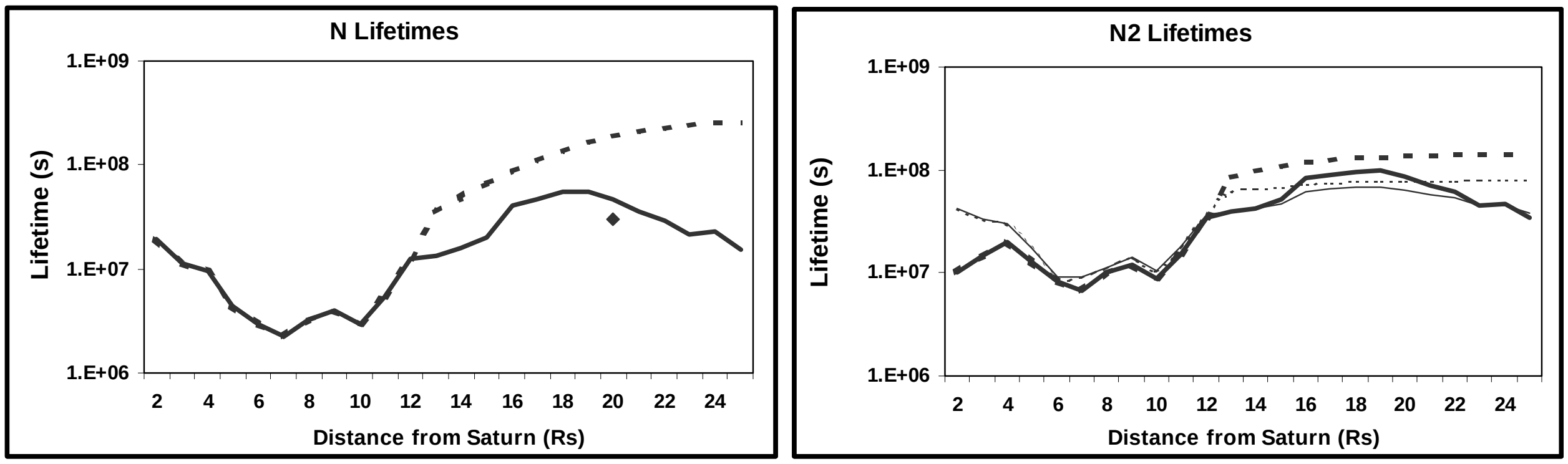

Figure 2. Total lifetimes for N (a) and $N_2$ (b) vs. distance from Saturn in Rs. Solid line: based on Richardson and Sittler (1990) and Voyager data, approximates periods when Titan is inside Saturn's magnetopause; Dotted line: beyond ~14Rs used a conservative extrapolation so that photo processes dominate at Titan's orbit, which would be the case when Titan is outside Saturn's magnetopause. Thick lines (b) for ionization and thin lines (b) for dissociation. Diamond: lifetime (a) used by Barbosa (1987).

**(a)** **(b)**

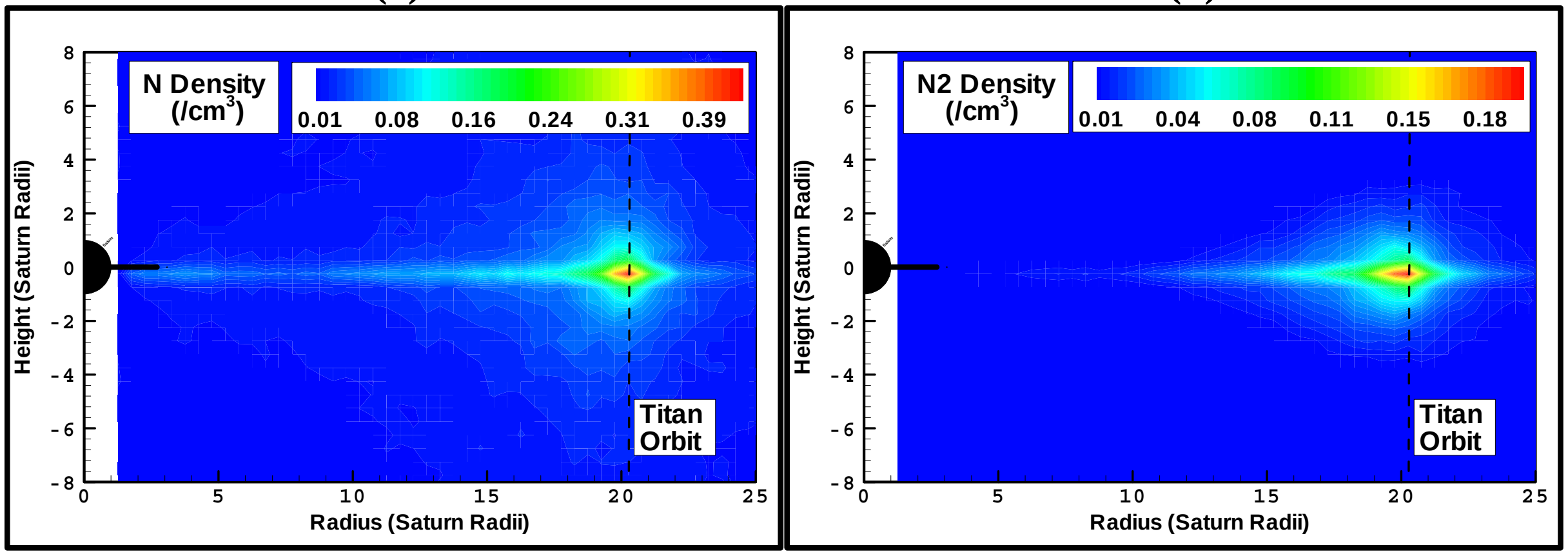

Figure 3 Atomic (a) and molecular (b) nitrogen densities using lifetimes from solid curves in Fig.2a and b. A 2-D vertical slice of the 3-dimensional torus in particles per $cm^3$. Saturn is on the left; position of Titan's orbit indicated by dashed line.

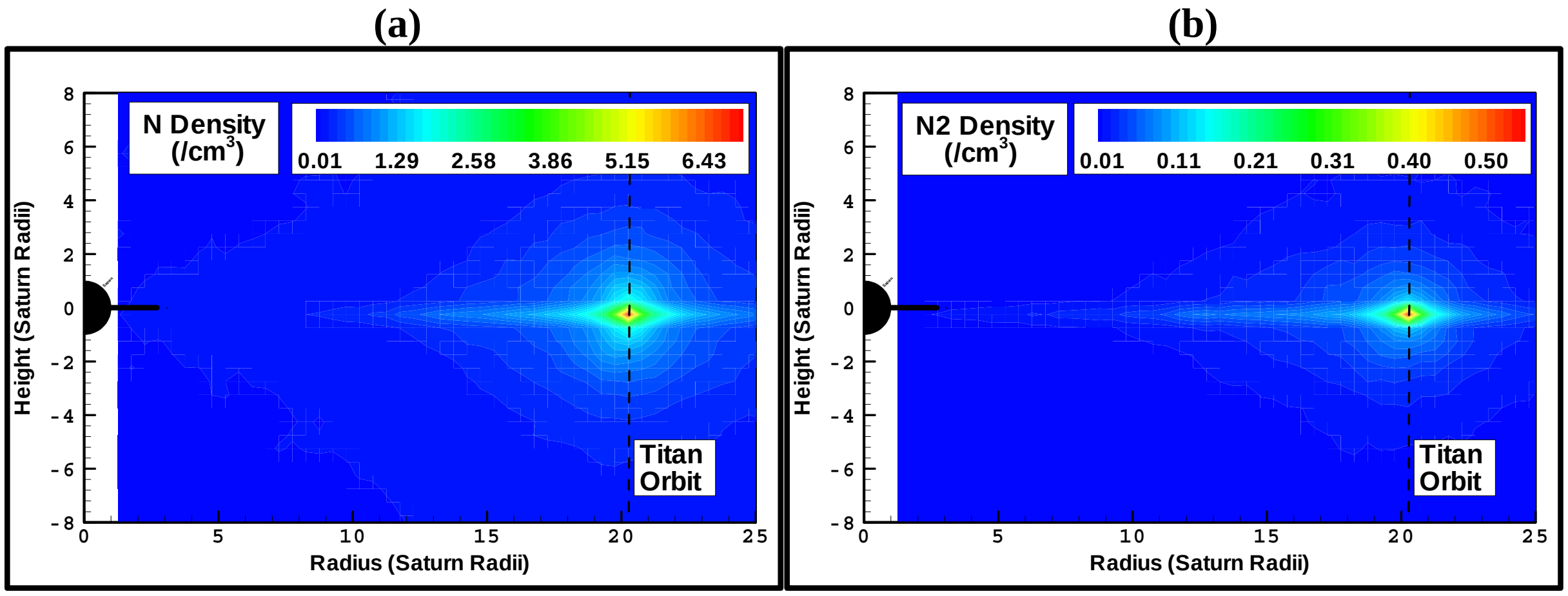

Figure 4. Atomic (a) and molecular (b) nitrogen densities generated when the conservative ionization and dissociation rates are used (dotted curves in Fig.2a and b): as in Fig.3

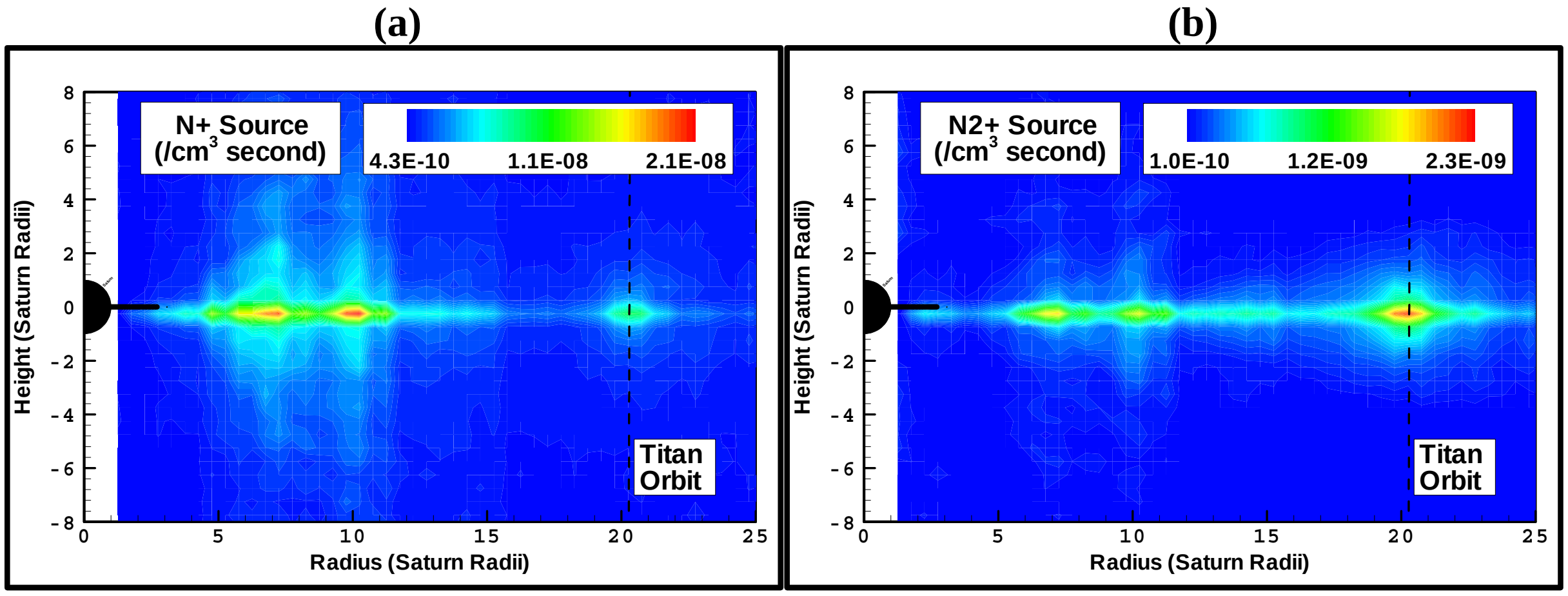

Figure 5. Computed N (a) and $N_2$ (b) ion source rates using lifetimes from solid curves in Fig.2a and b. A 2-D vertical slice of the 3-dimensional source distribution in particles per $cm^3$ per second. Saturn is on the left; position of Titan's orbit indicated by dashed line.